\documentclass{article}
\usepackage{spconf,amsmath,graphicx}
\usepackage{bm,amssymb,amsthm,amsfonts,algorithmic,mathrsfs,multirow,enumitem,caption,xcolor,mwe,stfloats,booktabs,textcomp}
\usepackage{fancyhdr}

\makeatletter
\let\NAT@parse\undefined
\makeatother
\definecolor{citecolor}{HTML}{0071BC}
\definecolor{linkcolor}{HTML}{FF6347}
\usepackage[pagebackref=true, colorlinks, citecolor=citecolor, linkcolor=linkcolor, bookmarks=false]{hyperref}

\def\tX{\mathcal{X}}

\def\t{\mathcal}

\definecolor{changecolor}{HTML}{38cb7d}

\title{Deep unrolling Shrinkage Network for Dynamic MR imaging}
\name{Yinghao Zhang$^*$, Xiaodi Li$^*$, Weihang Li$^\dagger$, Yue Hu$^*$\thanks{This work is supported by the National Natural Science Foundation of China under Grant 61871159 and Natural Science Foundation of Heilongjiang YQ2021F005.}}
\address{$^*$School of Electronics and Information Engineering, Harbin Institute of Technology, Harbin, China\\$^\dagger$School of Electrical Engineering and Automation, Tianjin University of Technology, Tianjin, China}
\begin{document}
%
\maketitle
\begin{abstract}
    Deep unrolling networks that utilize sparsity priors have achieved great success in dynamic magnetic resonance (MR) imaging. The convolutional neural network (CNN) is usually utilized to extract the transformed domain, and then the soft thresholding (ST) operator is applied to the CNN-transformed data to enforce the sparsity priors. However, the ST operator is usually constrained to be the same across all channels of the CNN-transformed data. In this paper, we propose a novel operator, called soft thresholding with channel attention (AST), that learns the threshold for each channel. In particular, we put forward a novel deep unrolling shrinkage network (DUS-Net) by unrolling the alternating direction method of multipliers (ADMM) for optimizing the transformed $l_1$ norm dynamic MR reconstruction model. Experimental results on an open-access dynamic cine MR dataset demonstrate that the proposed DUS-Net outperforms the state-of-the-art methods. The source code is available at \url{https://github.com/yhao-z/DUS-Net}.
\end{abstract}
%

\section{Introduction}
\label{sec:intro}

Dynamic MR imaging is widely used for non-invasive medical diagnoses, such as cardiac imaging and perfusion imaging. It captures and shows the temporal evolutions of objects of interest. However, the acquisition time of dynamic MR imaging is also substantially increased due to the additional introduction of the temporal dimension. 

Compressed Sensing \cite{lustig2008compressed, afonso2010fast} has been widely used to accelerate imaging by characterizing the sparsity structure of the data. The sparsity in the wavelet \cite{lustig2007sparse}, temporal Fourier \cite{tremoulheac2014dynamic} and total variation \cite{lingala2011accelerated} domain has been exploited and obtained promising results. Low-rank priors have also been explored in dynamic MR imaging. Typically, low-rank constraints are imposed on the Casorati matrices \cite{lingala2011accelerated, tremoulheac2014dynamic} to characterize the correlations between similar temporal evolutions of the same tissues. Recent works \cite{roohi2017multi, zhang2022opt} have further explored tensor low-rank priors based on various tensor decompositions. In addition, the combination of sparsity and low-rank priors has been proposed \cite{lingala2011accelerated,tremoulheac2014dynamic} to improve reconstruction performance. Nevertheless, these model-based methods are usually solved by iterative optimization algorithms, which are sensitive to the choice of parameters and the initialization of the model. The adjustment of the parameters needs to be performed manually, which is time-consuming and un-robust.
 
Deep unrolling network \cite{monga2021algorithm, aggarwal2018modl} mitigates the aforementioned shortages by combining the merits of both the model-based and deep-learning methods. It unfolds the iterative steps of the model-based optimization algorithm into the iteration modules, which are then stacked to form a supervised deep-learning framework. The parameters of the optimization algorithm are trained end-to-end adaptively. In addition, convolutional neural networks (CNNs) can be utilized to extract a transformed domain that the low-rank or sparsity priors are more suitable. The deep unrolling network has proven to be an effective solution for dynamic MR imaging. In recent works, low-rank priors have been incorporated into deep unrolling networks for this task, specifically through the application of singular value thresholding (SVT) on the unfolded Casorati matrices, as demonstrated in \cite{ke2021learned, huang2021deep}. Additionally, tensor low-rank properties have been exploited within a CNN-learned transformed domain in \cite{zhang2022net}. However, the gradient of the singular value decomposition of the SVT will be numerically unstable if the distance between any two singular values is close to zero \cite{wang2021robust}. It will be challenging for the gradient descent training process. Therefore, sparsity properties may be more suitable for deep unrolling networks without any gradient issues. Some methods have employed soft thresholding (ST) on the CNN-transformed data to exploit sparsity properties in the deep unrolling framework, such as ISTA-Net \cite{zhang2018ista} and SLR-Net \cite{ke2021learned}.
However, one limitation of these methods is that they only use one threshold of the ST, despite having $N_c$ channels of the CNN-transformed data. Additionally, the two CNN-learned transforms in their methods are constrained to be inverse, which will limit the ability of CNN to learn implicit properties beyond the sparsity if the threshold of ST is learned to be zero. 
In contrast, the DCCNN \cite{schlemper2017deep} and L+S-Net \cite{huang2021deep} take a different approach by directly utilizing the CNN and ReLU \cite{glorot2011relu} to implicitly exploit the properties of the dynamic MR data. However, it has been shown that the ST operator can be effective in improving the feature learning ability compared to ReLU when the information is highly damaged \cite{zhao2019deep}. Therefore, in highly undersampled dynamic MR imaging scenarios, it may be beneficial to incorporate the ST operator to improve the reconstruction performance.

\begin{figure}
\centering
\includegraphics[width=0.8\linewidth]{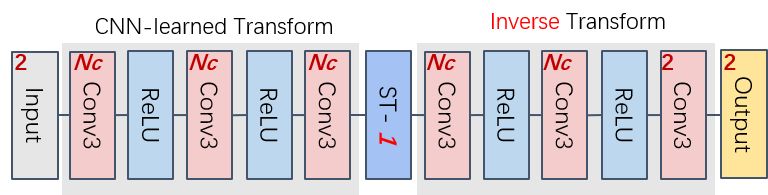}
\caption{The sparsity utilization of ISTA-Net and SLR-Net. The red number in the upper left corner of the box represents the number of channels.}
\label{fig:ISTA-SLR}    
\end{figure}

In this paper, we propose a deep unrolling shrinkage network for dynamic MR imaging, dobbed DUS-Net, which is inspired by the deep residual shrinkage network \cite{zhao2019deep}. A novel soft thresholding with channel attention \cite{hu2018squeeze} (AST) operator is proposed to adaptively learn the thresholds for each channel of the CNN-transformed data. The CNNs on both sides of AST are released without any constraints to learn the implicit properties of the dynamic MR data. In particular, the DUS-Net is unrolled from the alternating direction method of multipliers (ADMM) \cite{afonso2010fast} for optimizing a transformed $l_1$ norm dynamic MR reconstruction model. Experimental results on a cardiovascular cine MR dataset demonstrate the efficacy of the proposed DUS-Net.

\section{Proposed DUS-Net}
\label{sec:method}

In this section, we first briefly introduce the ADMM algorithm for optimizing a transformed $l_1$ norm dynamic MR reconstruction model and then present the proposed DUS-Net.

\subsection{ADMM for Transformed $l_1$ Norm Optimization}
\label{sec:admm}

The data acquisition of dynamic MR imaging can be modeled as
\begin{equation}
  \label{dmri_acquisition}
\mathbf b = \mathscr A(\mathcal{X}) +\mathbf{n},
\end{equation}
where $\mathcal{X} \in \mathbb{C}^{n_{x} \times n_{y} \times n_{t}}$ denotes distortion-free dynamic MR data, $n_x$, $n_y$ denote the spatial coordinates, $n_t$ is the temporal coordinate, $\mathbf b \in \mathbb{C}^{m}$ is the observed undersampled $k$-space data, $\mathscr A: \mathbb{C}^{n_{x} \times n_{y} \times n_{t}} \rightarrow \mathbb{C}^m$ is the Fourier sampling operator, and $\mathbf{n} \in \mathbb{C}^{m}$ is the Gaussian distributed white noise.

Then, the transformed $l_1$ norm dynamic MR reconstruction model can be formulated as
\begin{equation}
  \tX^* = \arg \min_{\tX} \frac 12\|\mathscr A(\tX) - \mathbf{b}\|_F^2 + \lambda \|\mathscr{T} (\tX)\|_{l_1},
\end{equation}
where $\|\mathscr A(\tX) - \mathbf{b}\|_F^2$ is the data fidelity that guarantees the consistency between the $k$-space of the reconstruction and the observation, the second term is the transformed $l_1$ regularization term, $\mathscr{T}$ denotes a unitary transform and $\lambda$ is the balancing parameter. 

By introducing an auxiliary variable $\t{Z}$, the optimization can be decoupled as,
\begin{equation}
    \min_{\tX} \frac 12 \Vert \mathscr A(\tX)-\mathbf{b} \Vert_F^2+\lambda \|\mathscr{T}(\mathcal{Z})\|_{l_1}
    \ \ s.t. \  \t{Z} =  \tX.
\end{equation}
The augmented Lagrangian function of the above optimization problem is formulated as,
\begin{equation}
  \begin{aligned}
      \label{augmentedlagrangian}
  \mathscr{L}(\tX, \t{Z}, \t{W}) = \frac 12 \Vert \mathscr A(\tX)-\mathbf{b} \Vert_F^2+\lambda \Vert \mathscr{T}(\mathcal{Z}) \Vert_{l_1} + \\
  <\t{W}, \tX - \t{Z}> + \frac {\mu}2 \Vert \t{Z} - \tX \Vert_F^2,
  \end{aligned}
\end{equation}
where $\t{W}$ is the Lagrangian multiplier and $\mu > 0$ is the penalty parameter. The ADMM \cite{afonso2010fast} is applied, leading to the following subproblems,
\begin{align}
    & \t{Z}_{n} = \min_{\t{Z}} \lambda \Vert \mathscr{T}(\mathcal{Z}) \Vert_{l_1} + \frac {\mu}2 \Vert \t{Z} - \tX_{n-1} - \t{L}_{n-1}\Vert_F^2, \label{z}\\
    & \tX_{n} = \min_{\tX} \frac 12 \Vert \mathscr A(\tX)-\mathbf{b} \Vert_F^2 + \frac {\mu}2 \Vert \t{Z}_{n} - \tX  - \t{L}_{n-1}\Vert_F^2, \label{x}\\
    & \t{L}_{n} = \t{L}_{n-1} - \eta (\t{Z}_{n}-\tX_{n}),
\end{align}
where $\t{L} = \frac {\t{W}} \mu$, the subscript $n$ denotes the $n$th iteration and $\eta$ denotes the update rate. 
The $\t{Z}$ subproblem can be solved by the soft thresholding operator \cite{donoho1995ST} since the transform $\mathscr{T}$ is defined as a unitary one \cite{ke2021learned}. The $\t{X}$ subproblem is quadratic and thus can be efficiently solved. Especially, when the samples are in the Cartesian grid, we can obtain a closed solution of the $\t{X}$ subproblem. Else, the conjugate gradient (CG) method can be used to solve it. Finally, the transformed $l_1$ norm dynamic MR reconstruction model obtains the following iterative steps,
\begin{equation}
    \small
    \label{iter_alg}
    \begin{cases}
          &\t{Z}_{n} = \mathscr{T}^H\circ\operatorname{ST}\left(\mathscr{T}(\tX_{n-1} + \t{L}_{n-1}), {\lambda}/{\mu}\right) \\
          &\tX_{n} = (\mathscr{A}^H \circ \mathscr{A} + \mu)^{-1}(\mathscr{A}^H(\mathbf{b})+\mu \t{Z}_{n} - \mu \t{L}_{n-1})\\
          &\t{L}_{n} = \t{L}_{n-1} - \eta (\t{Z}_{n}-\tX_{n})
    \end{cases}.
\end{equation}%
where $\circ$ represents the composition of operators, $\mathscr{T}^H$ denotes the adjoint of $\mathscr{T}$, $\operatorname{ST}$ denotes the soft thresholding operator and ${\lambda}/{\mu}$ is the threshold of ST.

\subsection{DUS-Net Framework}

\begin{figure}[htbp]
    \centering
    \includegraphics[scale=0.37]{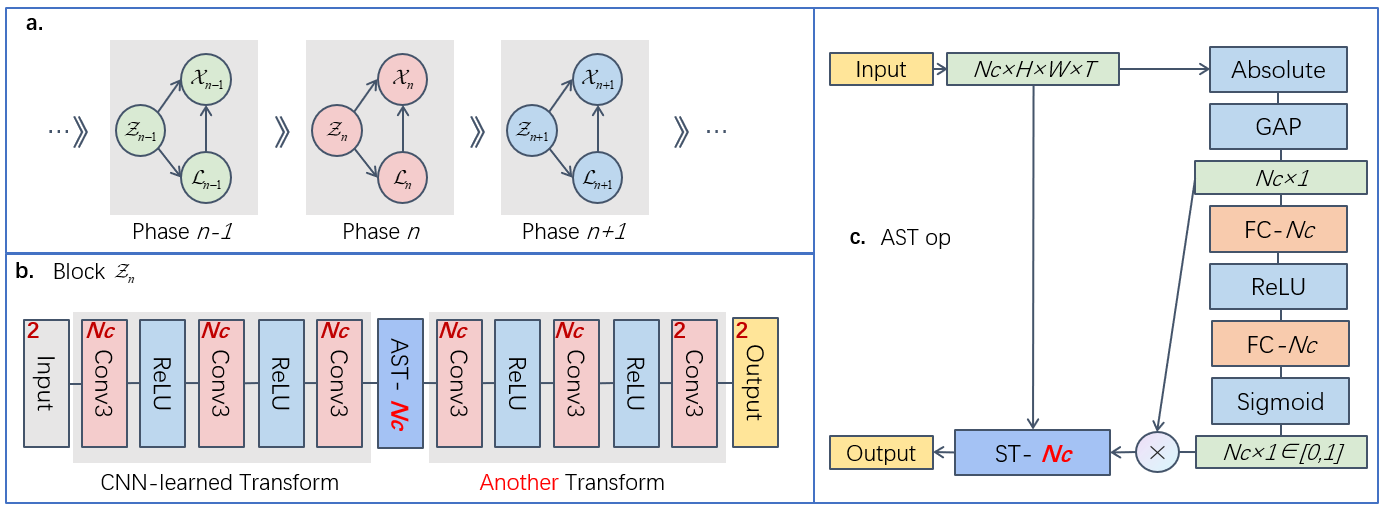}
    \caption{The framework of the proposed DUS-Net. a. displays the unrolling architecture of the DUS-Net with three adjacent phases shown, b. shows the $\mathcal{Z}_n$ block in detail, and c. shows the AST operator. The red number in the upper left corner of the box represents the number of channels.}
    \label{fig:framework}
\end{figure}

We establish the DUS-Net by unrolling the iterative ADMM procedure \eqref{iter_alg}. In particular, the transform $\mathscr{T}$ and its adjoint $\mathscr{T}^H$ are replaced by two separate CNNs to learn a suitable transformed domain for better utilization of sparsity. These two CNNs are released from the unitary constraint in \eqref{iter_alg} to characterize and utilize the implicit deep image priors. We propose a novel soft thresholding operator with channel attention (AST) to strengthen the traditional ST by assigning an independent learnable threshold for each channel.

The unrolling architecture of the DUS-Net is shown in Fig. \ref{fig:framework}. Each iterative step in \eqref{iter_alg} is unrolled into one phase of the DUS-Net. $N$ phases are then stacked to form the DUS-Net, where $N$ becomes a hyperparameter of the network. There are three fundamental blocks in each phase. The $\mathcal{Z}$ block corresponding to the $\mathcal{Z}$ step in \eqref{iter_alg} is related to the transformed sparsity utilization, while the other two blocks are exactly consistent with the $\tX$ and $\t{L}$ steps. 

In the $\mathcal{Z}$ block, the traditional ST is replaced by the novel AST, which is illustrated in Fig. \ref{fig:framework}(c). Specifically, given an input of size $N_c \times H \times W \times T$, where $N_c$ is the channel dimension and $H$, $W$, and $T$ are the spatial and temporal dimensions, the AST operator first takes the absolute value of the input. Global average pooling (GAP) \cite{goodfellow2016dl, lin2013network} is then applied to obtain a 1D vector of size $N_c \times 1$. This vector is fed into two fully connected (FC) layers \cite{goodfellow2016dl} separated by a ReLU \cite{glorot2011relu} activation. The output of the second FC layer is then passed through a sigmoid activation to obtain channel-wise attention weights \cite{hu2018squeeze}, which are constrained to the range $[0,1]$. The attention weights are multiplied by the output vector of the GAP, and the resulting channel-wise thresholds are obtained. The channel-wise ST is then applied, with the $i$-th threshold shrinking the $i$-th channel of the input. The AST operator is learned to adaptively assign thresholds for each channel and is differentiable, enabling end-to-end training of the DUS-Net.

In addition, two separate CNNs are utilized in each phase to feature the sparsity priors and other helpful implicit deep image priors. Unlike previous works that constrain these CNNs to be unitary or inverse \cite{zhang2018ista,ke2021learned}, we allow them to freely adapt and learn the priors. The reason is that if the AST operator is learned to be invalid (all pixel values are larger than the threshold) under the unitary or inverse constraint, the CNN-learned transform $\mathscr{T}$ and its adjoint $\mathscr{T}^H$ will collapse into the identity operator, leading to the failure of the DUS-Net to learn the implicit deep image priors via the CNN. Thus, the improvement of performance will be limited. Therefore, we choose to allow the CNNs to learn the priors in a flexible manner, and use only the Mean Square Error (MSE) as the loss function.

For implementation details, the learnable parameters include the weights of the CNNs, the weights of the AST operator and the parameters $\lambda$, $\mu$, $\eta$. 
To handle complex-valued data, we split each data point into two real-valued channels. The kernel size of the CNNs is set to 3*3*3, and most of the convolutional layers in each $\mathcal{Z}$ block have $N_c$ channels, except for the last layer which has only 2 channels to reconstruct the dynamic MR data. The FC layers in the AST operator have $N_c$ channels as well. In our implementation, we set $N_c$ to 16, and the number of phases in the DUS-Net is set to 15. To train the network, we adopt the exponential decay learning rate \cite{zeiler2012adadelta} with an initial learning rate of 0.001 and a decay of 0.95. The Adam optimization \cite{kingma2014adam} is adopted in the framework of TensorFlow \cite{abadi2016tensorflow}. The training is conducted on a single NVIDIA Tesla GV100 GPU with 32GB memory. 

\section{Experimental Results}
\label{sec:exp}

We evaluate the proposed DUS-Net using an open-access cardiovascular cine MR dataset OCMR \cite{ref_ocmr}. The dataset contains 204 fully sampled dynamic cine MR data from 3T Siemens MAGNETOM Prisma, 1.5T Siemens Avanto and 1.5T Siemens Sola machines. We use 124 for training, 40 for validation, and 40 for testing, in a ratio of about 7:2:2. We crop the training data into the size of 128$\times$128$\times$16 ($x\times y\times t$) for data augmentation, and the strides along the three dimensions are set as 32, 32 and 8, respectively. Finally, we obtain 1848 fully sampled training data. Note that all the multi-coil data from OCMR are combined into single-coil images, and the coil sensitivity maps are computed by ESPIRiT \cite{ref_multicoil}. 

\begin{figure*}[htbp]
    \vspace{-10 pt}
    \centering
    \includegraphics[width=0.85\linewidth]{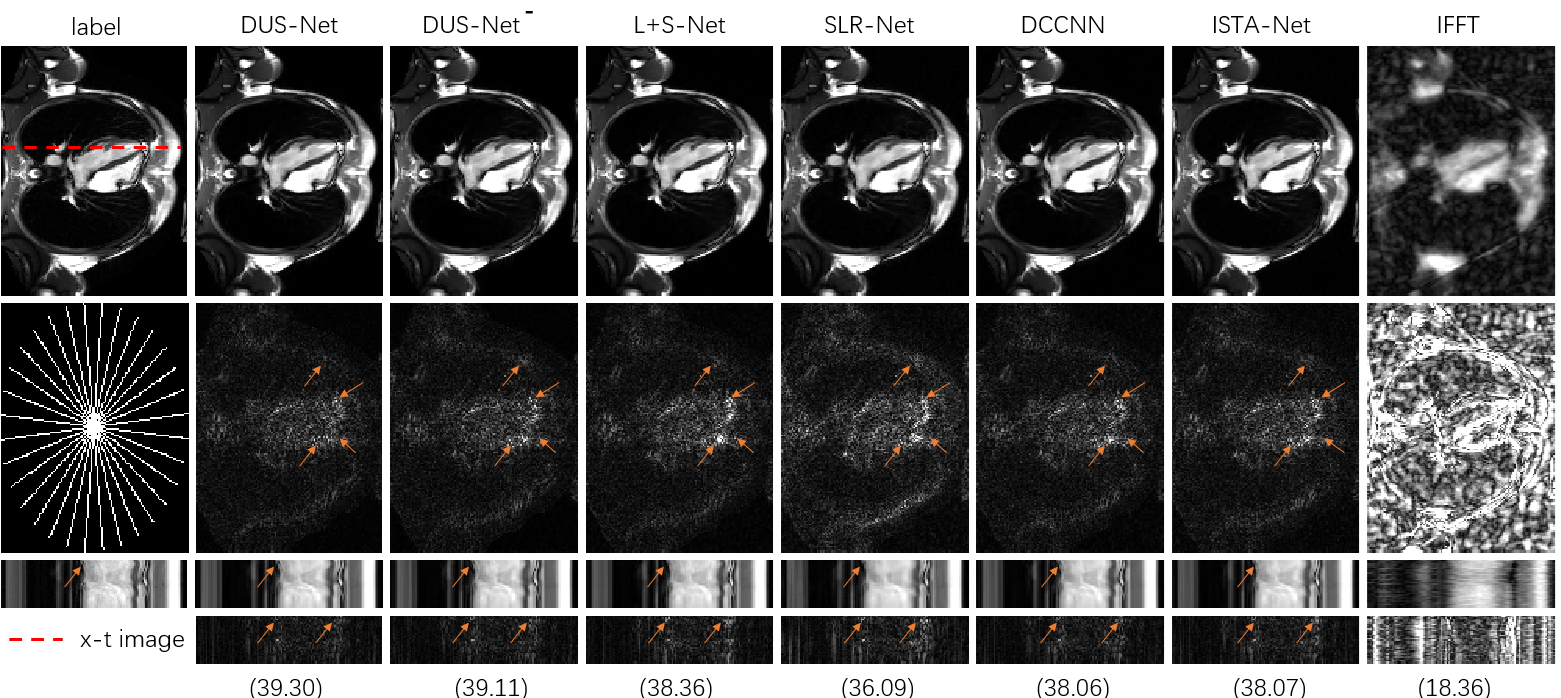}
    \caption{The reconstructed results of a specific test data under pseudo-radial sampling pattern \cite{lingala2011accelerated} with 16 spokes. The first row shows the reconstructed images, and the second row shows the corresponding error maps. The x-t slices and the corresponding error maps are shown in the third and fourth rows, respectively. The significant improvement of the DUS-Net is marked by the orange arrows. The reconstructed PSNRs are shown at the bottom.}
    \label{fig:result}
\end{figure*}

\begin{table}[htbp]
    \centering
    \small
    \caption{The quantitative comparison of the proposed DUS-Net with the state-of-the-art methods. The results are reported in the format of mean (standard deviation).}
    \resizebox{\linewidth}{!}{
    \begin{tabular}{lcccc}
    \bottomrule
        ~ & PSNR & SSIM & TIME/s & Parameters \\ \bottomrule
        \textbf{DUS-Net} & \textbf{41.617(2.582)} & \textbf{0.982(0.008)} & 0.80 & 448829 \\ 
        DUS-Net$^-$ & 41.428(2.592) & 0.980(0.008) & 0.83 & 440684 \\ \hline
        L+S-Net & 41.161(2.519) & 0.979(0.008) & 0.87 & 328339 \\ 
        ISTA-Net & 40.738(2.515) & 0.978(0.008) & 0.52 & 293780 \\ 
        DCCNN & 40.584(2.484) & 0.977(0.008) & 0.44 & 293760 \\ 
        SLR-Net & 38.655(2.408) & 0.964(0.010) & 0.91 & 293808 \\ \bottomrule
    \end{tabular}}
    \label{tab:exp}
\end{table}

\begin{table}[htbp]
    \centering
    \caption{The average reconstruction PSNRs of the DUS-Net under different sampling patterns. radial-30 denotes the pseudo-radial sampling pattern \cite{lingala2011accelerated} with 30 spokes. vds-8 and vds-10 denote the variable density sampling patterns with 8- and 10-fold acceleration, respectively. '-' denotes that the L+S-Net failed to converge due to the instability of SVD \cite{wang2021robust}.}
    \resizebox{\linewidth}{!}{
    \begin{tabular}{lccccc}
    \hline
        ~ & \textbf{DUS-Net} & L+S-Net & ISTA-Net & DCCNN & SLR-Net \\ \hline
        radial-30 & \textbf{43.29} & 43.01 & 42.96 & 42.90 & 40.21 \\
        vds-8 & \textbf{41.79} & - & 41.11 & 40.66 & 39.06 \\ 
        vds-10 & \textbf{34.88} & - & 34.79 & 34.60 & 33.88 \\ \hline
    \end{tabular}}
    \label{tab:exp2}
\end{table}

We compared the DUS-Net with four state-of-the-art (SOTA) deep unrolling methods, namely L+S-Net \cite{huang2021deep}, SLR-Net \cite{ke2021learned}, DCCNN \cite{schlemper2017deep}, and ISTA-Net \cite{zhang2018ista}. To evaluate the effectiveness of the AST operator, we also proposed a weakened version of DUS-Net, called DUS-Net$^-$, which is the same as the DUS-Net except that the AST operator is replaced by the traditional ST operator. All methods were trained for 50 epochs with a mini-batch size of 1 using the same training settings.
The reconstructed results from a specific data of the test dataset under the pseudo-radial sampling with 16 spokes are shown in Fig \ref{fig:result}. The comparison with four SOTA methods demonstrates that DUS-Net achieves superior performance in both qualitative visualization and quantitative PSNR. The orange arrows indicate that DUS-Net retains the finest tissue details and has the lowest reconstruction error. Table \ref{tab:exp} shows the quantitative results of DUS-Net and the SOTA methods under pseudo-radial sampling pattern \cite{lingala2011accelerated} with 16 spokes, evaluated by the PSNR and SSIM. The number of parameters and the average inference time are also reported. DUS-Net outperforms the SOTA methods in both PSNR and SSIM, with improvements of approximately 0.5dB and 1dB compared to the second-best and other SOTA methods, respectively.
The DUS-Net outperforms the DUS-Net$^-$, demonstrating the effectiveness of the AST operator. The inference time of DUS-Net is comparable to other SOTA methods, with improved accuracy. Surprisingly, the DUS-Net is faster than the DUS-Net$^-$, as the additional FC layers in AST do not significantly increase the inference time. This may be due to the parallel computation of the channel-wise operation in AST. However, in our implementation, the SLR-Net has poorer performance than the ISTA-Net, which may be due to the stricter inverse constraint of SLR-Net on the two CNN-learned transforms. Moreover, we investigated the robustness of the DUS-Net under different sampling patterns and present the results in Table \ref{tab:exp2}.

To demonstrate the effectiveness of releasing the inverse constraint of the two CNNs, we have included it as part of the loss function, along with a balancing hyperparameter $\zeta$, similar to ISTA-Net and SLR-Net. The inverse constraint is formulated as $\sum_{i=1}^{N}\|\widehat{\mathscr{F}} \circ \mathscr{F} (\tX_i)-\tX_i\|_F^2$, where $\widehat{\mathscr{F}}$ and $\mathscr{F}$ denote the two CNNs in each phase, and $N$ is the total phase number. We investigated the average reconstruction PSNR of the DUS-Net for three different values of $\zeta$: 0.001, 0.01, and 0.1, under the same sampling pattern as Table \ref{tab:exp}. The resulting PSNR values were 41.59, 40.95, and 37.00 dB, respectively. Compared with the higher PSNR value of 41.62 dB without constraint in Table \ref{tab:exp}, it can be concluded that releasing the inverse constraint leads to better reconstruction performance.

\section{Conclusion}
\label{sec:conclusion}
\vspace{-2 pt}
In this work, we introduce a deep unrolling shrinkage network (DUS-Net) for dynamic MR imaging, which is unrolled from the ADMM algorithm to optimize the transformed $l_1$ norm dynamic MR model. Our approach employs two CNNs in each unrolling phase to adaptively extract a suitable transformed sparsity domain and a novel soft thresholding with channel attention (AST) operator to adaptively shrink the CNN-transformed data with the channel-wise thresholds. We also release the inverse constraint between the two CNNs, allowing us to utilize beneficial implicit deep image priors. The experimental results on the OCMR dataset demonstrate that the DUS-Net outperforms state-of-the-art unrolling methods in both qualitative and quantitative performance.



\clearpage

\bibliographystyle{IEEEbib}
\small\bibliography{refs}

\end{document}